\documentclass{PoS}
\usepackage{url}
\usepackage{amsmath}

\title{Numerical Implementation of Gauge-Fixed Fourier Acceleration}

\ShortTitle{Numerical Implementation of Gauge-Fixed Fourier Acceleration}

\author{\speaker{Yidi Zhao}\thanks{This work was partially supported by US DOE grant \#DE-SC0011941 and the DOE Exascale Computing Project (17-SC-20-SC), a collaborative effort of the U.S. Department of Energy Office of Science and the National Nuclear Security Administration.}\\
        Columbia University\\
        E-mail: \email{yz3210@columbia.edu}}

\abstract{In hybrid Monte Carlo evolution, by imposing a physical gauge condition, simple Fourier acceleration can be used to generate conjugate momenta and potentially reduce critical slowing down. This modified gauge evolution algorithm does not change the gauge-independent properties of the resulting gauge field configurations. We describe this algorithm and it numerical implementation.}

\FullConference{The 36th Annual International Symposium on Lattice Field Theory - LATTICE2018\\
		22-28 July, 2018\\
		Michigan State University, East Lansing, Michigan, USA.}

\begin{document}

\section{Introduction}

In the hybrid Monte Carlo (HMC) evolution of gauge configurations, as the spacing between lattice points is decreased, the gauge fields can be separated into short-distance, stiff modes, as well as relatively long-distance and less constrained modes. The conventional HMC algorithm does not differentiate these modes; long-distance and short distance modes evolve with the same velocity. The molecular dynamics step size is limited by the short-distance modes and is smaller than what is necessary for the long-distance modes, while the number of steps is restricted by the long-distance modes and is more than what is necessary for the short-distance modes. This leads to slowing down of gauge configuration evolution and is usually referred to as critical slowing down.
 
We present an algorithm that is intended to apply Fourier acceleration to reduce critical slowing down and accelerate the generation of gauge field ensembles. In contrast to the conventional HMC algorithm, we use a different fictitious mass for different modes, arranged so that the velocities for the long-distance modes will be much larger than those of the short-distance modes. One barrier to applying Fourier acceleration is that for the QCD action, the spatial dependence of the different modes of the gauge field is obscured by Yang-Mills gauge symmetry. This problem can be overcome if we introduce some sort of physical gauge fixing. In our implementation, we introduce a gauge-fixing term into the action to softly fix the gauge fields to Landau gauge. Instead of fixing the gauge rigidly, the gauge fields are allowed to move in a small neighborhood of Landau gauge.

\section{Gauge Fixing}

To resolve the difficulty resulting from gauge symmetry, we introduce the following term into the action:

\begin{equation}
S_{GF1}[U] = - \beta M^2 \sum_{x,\mu} Re\ tr[U_\mu(x)].
\end{equation}

When this action is at minimum, the gauge links are constrained to Landau gauge without Gribov copies \cite{Zwanziger90, Parrinello90}. When evaluating the path integral with this additional term, gauge field configurations that are close to Landau gauge get more weight, while gauge field configurations farther away from Landau gauge are suppressed. $M$ appears as a mass-like parameter which can be tuned to determine how rigidly Landau gauge fixing is imposed. If we choose $M$ to be a large number then the gauge links will be effectively restricted to a small neighborhood of Landau gauge.

The behavior of this term in the weak coupling limit has been analyzed through one loop in Ref. \cite{Serreau:2013ila}. Compared to other possible gauge-fixing actions, this one is easier to implement and is relatively computationally cheap. Its linear dependence on each link allows the use of the usual heatbath algorithm to generate gauge transformation fields in an inner Monte Carlo which will be discussed soon.

After the introduction of this gauge-fixing term, another compensating term is needed to ensure that gauge-fixing does not affect any gauge-independent properties of the resulting gauge configurations. Using the gauge invariance of the original pure gauge action and properties of the Haar measure, one finds that for any gauge-invariant observable $O$:

\begin{align}
\langle O\rangle &= \int dU e^{-S[U]} O[U] \\
&= \int dU \frac{\int dg^\prime e^{-S_{GF1}[U^{g^\prime}]}e^{-S[U]} O[U]}{\int dg e^{-S_{GF1}[U^g]}} \\
&= \int dU \frac{\int dg^\prime e^{-S_{GF1}[U^{g^\prime}]}e^{-S[U^{g^\prime}]} O[U^{g^\prime}]}{\int dg e^{-S_{GF1}[U^g]}} \\
& = \int dU e^{ -S[U] -S_{GF1}[U] -\ln \int dg e^{-S_{GF1}[U^g]}} O[U],
\end{align}
where $U^g$ represents the gauge configuration after gauge transformation, i.e. $U^g_\mu(x) = g(x) U_\mu(x) g(x+\hat{\mu})^\dagger$. Thus we can add the following compensating term to the action

\begin{equation}
S_{GF2}[U] = \ln \int dg\ e^{-S_{GF1}[U^g]}.
\end{equation}

Adding $S_{GF1}$ and $S_{GF2}$ into action will not change the expected value of any gauge-invariant operator. The resulting Green's functions of gauge-invariant quantities will be identical to those generated with the original gauge action. To sum up, our total action is:

\begin{align}
S[U] &= S_{wilson}[U] + S_{GF1}[U] + S_{GF2}[U] \\
S_{GF1}[U] &= -\beta M^2 \sum_{x,\mu} Re\ tr[U_\mu(x)] \\
S_{GF2}[U] &= \ln \int dg\ e^{-S_{GF1}[U^g]}.
\end{align}

The molecular dynamics force, which acts on the conjugate momenta in the hybrid Monte Carlo integration, is calculated from the derivative of action. The compensating term $S_{GF2}$ involves an integral over gauge transformation fields, which brings in major difficulties for this algorithm. One approach to solve this difficulty is to make use of importance sampling and replace the integral with a sum over gauge transformations distributed according to $e^{-S_{GF1}[U^g]}$.

\begin{equation}
F_l = \frac{\partial}{\partial U_l} \{ S_{wilson}[U] + S_{GF1}[U] \} - \sum_{n=1}^{N}\frac{\partial S_{GF1}[U^{g^n}]}{\partial U_l}.
\end{equation}

Then calculating the molecular dynamics force stemming from $S_{GF2}$ becomes a standard Monte Carlo computation. Similarly, in the accept-reject step of the HMC algorithm, we are supposed to compute the change in the action between the beginning and the end of the a trajectory. $S_{GF2}$ cannot be calculated directly; however, the difference in $S_{GF2}$ is calculable

\begin{align}
S_{GF2}[U^\prime] - S_{GF2}[U] &= \ln \frac{\int dg\  e^{-S_{GF1}[U^{^\prime g}]}}{\int dg\  e^{-S_{GF1}[U^g]}} \\
&= \ln \frac{\int dg\ e^{-S_{GF1}[U^g]}  e^{S_{GF1}[U^g]-S_{GF1}[U^{^\prime g}]}}{\int dg\  e^{-S_{GF1}[U^g]}} \\
&\approx \ln \frac{1}{N} \sum_{n=1}^{N} e^{S_{GF1}[U^{g^n}]-S_{GF1}[U^{^\prime g^n}]}.
\end{align}

Thus, calculating $\Delta H$ also involves an average over gauge transformations chosen with the Monte Carlo weight $e^{-S_{GF1}[U^g]}$. This summation is a sum of exponentials and will be strongly dominated by a few exponentials which are extraordinarily large compared to the typical value. Therefore, computing the difference in the action is a challenge and may require a great number of samples to get an accurate value. 

\section{Fourier Acceleration}

As the mass parameter $M$ in the gauge fixing action is increased, the resulting configurations will be more nearly in Landau gauge and introducing Fourier acceleration into the molecular dynamics kinetic energy should be effective in decreasing critical slowing down.

In continuum limit, the effects of gauge-fixing action has been analyzed in Ref. \cite{Serreau:2013ila}. On the lattice, to achieve Fourier acceleration, we propose to use the following kinetic energy term

\begin{align}
H_p &= \sum_k tr(P_\mu(-k) D^{\mu\nu}(k) P_\nu(k)) \\
D_{\mu\nu}(k) &= \frac{1}{\sin(\frac{k}{2})^2+\epsilon^2} P^T_{\mu\nu}(k) + \frac{1}{M^2} P^L_{\mu\nu}(k) \\
P^T_{\mu\nu}(k) &= \delta_{\mu\nu} - \frac{e^{-i\frac{k_\mu}{2}} \sin(\frac{k_\mu}{2})   e^{i\frac{k_\nu}{2}} \sin(\frac{k_\nu}{2})}{\sin(\frac{k}{2})^2} \\
P^L_{\mu\nu}(k) &=  \frac{e^{-i\frac{k_\mu}{2}}  \sin(\frac{k_\mu}{2})   e^{i\frac{k_\nu}{2}} \sin(\frac{k_\nu}{2})} {\sin(\frac{k}{2})^2}.
\end{align}

As required for Fourier acceleration, this choice is the inverse of the quadratic part of our entire gauge-fixed action \cite{Serreau:2013ila}. Since $P^T$ and $P^L$ appearing in the equation above are projection operators, the square root of the coefficient matrix $D_{\mu\nu}$ in that equation can be readily evaluated and thus conjugate momenta $P_\nu(k)$ can be generated by selecting samples from a Gaussian distribution and multipying them by $\sqrt{D}_{\mu\nu}$.

\section{Conclusion}

In this article, we have described a new algorithm aimed at reducing critical slowing down of the HMC algorithm for lattice QCD. Several components must be combined to realize soft Landau gauge fixing, which  introduce an inner Monte Carlo that is computationally expensive and brings in statistical errors. Numerical experiments are required to test the feasibility of this algorithm and to search for proper parameters. Especially, the extra computational workload caused by inner Monte Carlo needs to be determined and optimized; also, we are supposed to choose an appropriate soft gauge-fixing parameter $M$ which can effectively reduce critical slowing down and at the same time avoid bringing in too large a statistical error. Code has been written based on the Grid system and is now being tested.

The gauge-fixed Fourier acceleration algorithm affects only gauge evolution and will work equally well for any type of dynamical fermions. Thus further numerical experiments are required to examine how effectively it can reduce critical slowing down with dynamical fermions. Inner Monte Carlo makes the pure gauge evolution much more computationally costly, but numerical tests are needed to compare its computational complexity with that of dynamical fermions and to determine if this algorithm can make the overall evolution more efficient.

\bibliographystyle{JHEP}
\bibliography{GFFA}

\end{document}